\documentclass[%twocolumn,
aps,nofootinbib,showpacs,showkeys,preprint
tightenlines,preprintnumbers,] {revtex4}

\usepackage{epsf,epsfig,subfigure,graphicx,amsmath,amssymb}
\usepackage{color}
\newcommand{\dis}[1]{\begin{equation}\begin{split}#1\end{split}\end{equation}}

\newcommand{\gev}{\,\textrm{GeV}}

\newcommand{\GSM}{{\rm SU(3)}_c\times {\rm SU(2)}_W\times{\rm U(1)}_Y}
\newcommand{\sMSSM}{$\sigma$MSSM}
\newcommand{\Mg}{{M_{\rm GUT}}}
\newcommand{\ie}{{\it i.e.}\ }
\newcommand{\etal}{{\it et al.}}
\newcommand{\Z}{{\bf Z}}

\def\E6{{\rm E_6}}
\def\EE8{{\rm E_8\times E_8}}

\begin{document}

\title{Couplings  between QCD axion and photon from string compactification}

\author{Jihn E.  Kim}
\address
{Center for Axion and Precision Physics Research (IBS),
  291 Daehakro,  Daejeon 34141, Republic of Korea, and\\
Department of Physics, Kyung Hee University,\\
 26 Gyungheedaero,  Seoul 02447, Republic of Korea, and\\
Department of Physics, Seoul National University, 
 1 Gwanakro,  Seoul 08826, Republic of Korea 
}
\begin{abstract}
The QCD axion couplings of various invisible axion models are presented. In particular,   the exact global symmetry U(1)$_{\rm PQ}$  in the superpotential is possible for the anomalous U(1) from string compactification,  broken only by the gauge anomalies at one loop level, and is shown to have the resultant invisible axion coupling to photon,   $c_{a\gamma\gamma}\ge \frac83-c_{a\gamma\gamma}^{\rm ch\,br}$ where $c_{a\gamma\gamma}^{\rm ch\,br}\simeq 2$. However, this bound is not applicable in  approximate U(1)$_{\rm PQ}$ models  with sufficiently suppressed U(1)$_{\rm PQ}$-breaking  superpotential terms. We also present a simple method to obtain $c_{a\gamma\gamma}^0$ which is the value obtained above the electroweak scale.

\keywords{Axion-photon-photon coupling,   KSVZ axion, DFSZ axion, Superstring axions}
\end{abstract}

\pacs{14.80.Va, 12.10.Kt, 11.25.Wx, 11.30.Fs}

\maketitle

%%%%%%%%%%%%%%
\section{Introduction}
       
The detection possibility of the invisible axions \cite{KSVZ1,KSVZ2,DFSZ} chiefly relies on its appreciable couplings to photon $c_{a\gamma\gamma}$ which appears in the Lagrangian as
\dis{
{\cal L}_{\rm axion\,coupling} =-\frac{a}{32\pi^2\,f_a}\left(c_3g_3^2G^a\tilde{G}^a+
c_{a\gamma\gamma}e^2F_{\rm em}\tilde{F}_{\rm em} \right),\label{eq:Defc3cem}
}
where
\dis{
G^a\tilde{G}^a = \frac{1}{2} \epsilon^{\mu\nu\rho\sigma}G^a_{\mu\nu} G^a_{\rho\sigma} ,~
F_{\rm em}\tilde{F}_{\rm em} = \frac{1}{2}\epsilon_{\mu\nu\rho\sigma}F_{\rm em}^{\mu\nu}\tilde{F}_{\rm em}^{\rho\sigma} .
}
The axionic domain-wall number \cite{SikivieDW} is $|c_3+c_2|$ where $c_2$ is the contribution from the standard model quarks \cite{KimRMP10}. With this normalization from the QCD sector,
the coupling $c_{a\gamma\gamma}$ is defined and is composed of two parts,
\dis{
c_{a\gamma\gamma}=c_{a\gamma\gamma}^0+ 
c_{a\gamma\gamma}^{\rm ch\, br}\simeq~ c_{a\gamma\gamma}^0-2
}
where  $c_{a\gamma\gamma}^0$  is the one obtained above the electroweak scale and $c_{a\gamma\gamma}^{\rm ch\, br}$  is the contribution obtained below the QCD chiral phase transition scale. Since the mass ratio of   up and down quarks is very close to 0.5 \cite{ManoharPDG}, we use the value $m_u/m_d=0.5$ below. In this case, $c_{a\gamma\gamma}^{\rm ch\, br}$ is --2 (a bit smaller value --1.98, including the strange quark contribution) \cite{KimPRP87}.  The early summary on the axion--photon--photon couplings were summarized in \cite{KimPRD98,KimRMP10}.\footnote{The earlier unification value was given in  \cite{Kaplan85}.}

%%%%%%%%%%%%%%%%%%%%
\begin{figure}[!t]
\begin{center}
\includegraphics[width=0.35\linewidth]{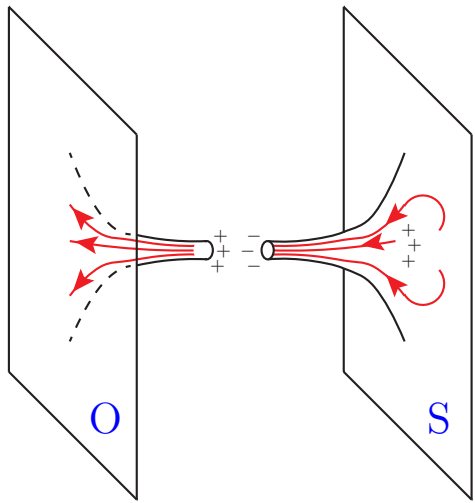}
\end{center}
\caption{Escaping charges through a wormhole.} \label{fig:Wormhole}
\end{figure}
%%%%%%%%%%%%%%%%%

An invisible axion is a pseudo-Goldstone boson whose mother symmetry is the Peccei-Quinn(PQ) symmetry \cite{PQ77}. The symmetry breaking scale relevant for the axion detection experiments \cite{ADMX,BCM14} are the intermediate scale $ 10^{10\,}\gev\lesssim f_a\lesssim 10^{12\,}\gev$, which can be achieved by the vacuum expectation value(VEV) of an SU(2)$\times$U(1)$_Y$ singlet $\sigma$ \cite{KSVZ1}. But, the global symmetry which is broken at the intermediate scale has to be fine-tuned to avoid the gravity spoil of global symmetries. This dificulty has been appreciated \cite{BarrGr92} after realizing that even the classical gravity does not necessarily preserve global symmetries due to the topology change via wormholes \cite{Giddings87,Coleman88}. The wormhole taking out gauge charged is depicted in Fig. \ref{fig:Wormhole}. An observer in the almost flat space O notices that some gauge charges are flown to the shadow world S.  To see the effect in his own space only, he cuts the wormhole, then notices that  the escaped gauge charges are recovered to O. This conservation of gauge charges in the space O is due to the long range electrix flux lines. For the global charges, there is no such flux lines and hence the escaped global charges are not considered to be recovered to O if he cuts the wormhole. Thus, global charges are broken if we consider the topology change.
Related to this effect, at field theory level within supersymmetric(SUSY) framework,
a host of discrete symmetries are considered \cite{StroWitten85, DiscrGauge89, Ibanez92, BanksDine92, Preskill91}. Some discrete symmetries can lead to acceptable approximate global symmetries \cite{KimPRL13,KimPLB13}.

In this paper, we attempt to obtain a region of the parameter space from string compactified 4-dimensional(4D) effective field theory. The 4D models from string compactification do not allow global symmetries but allow some discrete symmetries \cite{Kobayashi07}. The minimal supersymmetric standard model supplied with singlets $\sigma$ (to house the invisible axion as pointed out in \cite{KSVZ1}) will be called  \sMSSM. In the \sMSSM, we calculate the couplings between axion and photon.

%%%%%%%%%%%%%%
\section{Gauge transformation of model-independent axion}

Four dimensional pseudoscalars in the \sMSSM~  from string compactification appear from $B_{MN}  $(=10D antisymmetric tensor field with $M,N\in\{1,2,\cdots, 10\}$) and from the matter super-multiplets in the \sMSSM. In 10D,  $B_{MN}$ is a gauge field satisfying the gauge transformation,
\dis{
 B_{MN}\to  B_{MN} -\partial_M\Lambda_N+\partial_N\Lambda_M,
 \label{eq:BMNgauge}
}
where $\Lambda_M$ are  gauge functions. If both $M$ and $N$ take the internal space coordinates $i, j=\{5,6,\cdots,10\}$, $B_{ij}$ is a pseudoscalar in 4D. From the 4D point of view, the original gauge transformation is not local, not carrying the 4D index $\mu=\{1,2,3,4\}$. These pseudoscalars are the model-dependent (MD) axion \cite{WittenMD}, which is known to generate superpotential terms \cite{WenWitten}. So, the MD axions are not the useful candidates for solutions of the strong CP problem. On the other hand the model-independent(MI) axion \cite{GreenSch84,WittenMI}, where both $M$ and $N$ of $B_{MN}$ take 4D indices $\mu,\nu$, is a good candidate of 4D gauge transformation. Namely, the gauge transformation (\ref{eq:BMNgauge}) is still a gauge transformation in 4D. So, $B_{\mu\nu}$ is not spoiled by gravity after the compactification.

%%%%%%%%%%%%%%%%%%%
\subsection{Anomaly-free  U(1) gauge symmetry}

If the MI axion is not behaving as a longitudinal degree of a gauge boson, then the axion potential is generated and the bosonic collective motion behaves as cold dark matter (CDM) \cite{BCM14}. 
But, then the axion decay constant is near the string scale, $f_a>10^{16\,}\gev$
\cite{ChoiKim85}, and  a fine-tuning is needed, or the anthropic scenario must be invoked \cite{Pi84}. The coupling $c_{a\gamma\gamma}$ is the same as the one considered in the following subsection with anomalous U(1) without extra charged singlets.

%%%%%%%%%%%%%%%%%%%
\subsection{Anomalous U(1) gauge symmetry}

If the compactification produces an anomalous U(1)$_{\rm anom}$ gauge symmetry, the corresponding U(1)$_{\rm anom}$  gauge boson obtains a large mass, at a slightly lower scale than the string scale. The presence of  a Fayet-Illiopoulos D-term connects the MI axion with the anomalous U(1)$_{\rm anom}$  gauge boson \cite{AnomUone}, which is a kind of the Higgs mechanism providing the longitudinal degree of the gauge boson. The generator of the anomalous U(1)$_{\rm anom}$  belongs to the $\EE8$ algebra, and matter fields have the U(1)$_{\rm anom}$  charges. The field $B_{\mu\nu}$ or the MI axion does not couple to matter fields. So below the U(1)$_{\rm anom}$ gauge boson mass scale, the U(1)$_{\rm anom}$ charge of matter fields becomes a global charge which can be called a  U(1)$_{\rm PQ}$ charge. In this way, a global symmetry free of the gravity obstruction is created below the U(1)$_{\rm anom}$ gauge boson mass scale. Since the mother U(1)$_{\rm anom}$ is a gauge symmetry, there is no gravity obstruction of this U(1)$_{\rm PQ}$ global symmetry. In string compactification, it has been explicitly shown that the anomalies are the same for all gauge groups both for non-Abelian and (properly normalized) Abelian gauge fields \cite{KimPLB88,KimPLB14}.
Thus, the compactification of anomalous U(1) gauge symmetries can be gates to low energy gravity-safe global symmetries. In the compactification of Type-I and Type-IIB string, there can be three anomalous U(1)'s
\cite{Uranga99}, where however the full phenomenologically acceptable \sMSSM~spectrum of matter fields are not presented, which forbids us from calculating $c_{a\gamma\gamma}$. Here, we restrict to the case of $B_{\mu\nu}$ from the heterotic string. 

{ 
Let the anomalous gauge symmetry be U(1)$_{\rm anom}$. Its charge operator and the coupling constant be $\Gamma$ and $e_\Gamma$, respectively. The potential which is invariant under U(1)$_{\rm anom}$ is also invariant under the global symmetry U(1)$_\Gamma$ whose charge generator is also $\Gamma$. To see the effective global symmetry below the anomalous scale, therefore, it is sufficient to see how the local transformation is described. Since the longitudinal degree of the U(1)$_{\rm anom}$ gauge boson is solely provided by $B_{\mu\nu}$, matter scalars having the nonvanishing $\Gamma$ charge do not develop VEVs. To see the U(1)$_{\rm anom}$ gauge transformation of a complex scalar $\Phi$, consider the kinetic energy term $(D^\mu\Phi)^*(D_\mu\Phi)$ where $D_\mu=\partial_\mu-ie_\Gamma\Gamma A_\mu$. The gauge transformation $\Phi\to e^{i\alpha(x)}\Phi$ leads the kinetic energy term to
\dis{
&(\partial^\mu\Phi^* +ie_\Gamma \Gamma A^\mu \Phi^*)(\partial_\mu\Phi -ie_\Gamma \Gamma A_\mu \Phi) +(e^{i\alpha}\partial^\mu e^{-i\alpha})\Phi^*(\partial_\mu\Phi -ie_\Gamma \Gamma A_\mu \Phi)\\
&+(e^{-i\alpha}\partial_\mu e^{i\alpha}) (\partial^\mu\Phi^* +ie_\Gamma \Gamma A^\mu \Phi^*)\Phi
+(\partial^\mu e^{-i\alpha})(\partial_\mu e^{i\alpha})\Phi^*\Phi.
}
If we consider the global transformation U(1)$_\Gamma$ below the anomalous scale, only the first term survives in the above equation,
\dis{
\textrm{U(1)}_\Gamma\,\Phi:\quad  (\partial^\mu\Phi^* +ie_\Gamma \Gamma \tilde{Z}^\mu \Phi^*)(\partial_\mu\Phi -ie_\Gamma \Gamma \tilde{Z}_\mu \Phi) 
}
where we expressed the  U(1)$_{\rm anom}$ gauge boson as $\tilde{Z}$. Below the anomalous scale, it describes a global symmetry U(1)$_\Gamma$ coupling with the heavy anomalous gauge boson with the same charge $\Gamma$. In the potential $V$, this gauge boson coupling respects the global symmetry also. Thus, we obtain an exact global symmetry U(1)$_\Gamma$ below the anomalous scale. 

}

Thus,  an intermediate scale global symmetry is from the anomalous U(1)$_{\rm anom}$ in the compactification process of the heterotic string. Since the anomalous U(1)$_{\rm anom}$ has the same coupling to gauge fields, we have the following MI axion coupling,
 \dis{
{\cal L} &= \frac{ P}{f}\,\frac{g_2^2}{32\,\pi^2}W^{a}_{\mu\nu}\tilde{W}^{{a}\,\mu\nu} + \frac{P}{f}\,\frac{g_1^{2}}{32\,\pi^2}
Y_{1,\mu\nu}\tilde{Y}_1^{\mu\nu}= \frac{ P}{f}\,\frac{g_2^2}{32\,\pi^2}\left( W^{a}_{\mu\nu}\tilde{W}^{{a}\,\mu\nu} +   (1/C^2) 
Y_{1,\mu\nu}\tilde{Y}_1^{\mu\nu}\right)\\
&= \frac{ P}{f}\,\frac{g_2^2}{32\,\pi^2}\left( W^{a}_{\mu\nu}\tilde{W}^{{a}\,\mu\nu} +   Y_{\mu\nu}\tilde{Y}^{\mu\nu}\right)\to \frac{ P}{f}\,\frac{g_2^2}{32\,\pi^2}\left(2 W^{+}_{\mu\nu}\tilde{W}^{{-}\,\mu\nu} +  F^{\textrm{em}}_{\mu\nu}\tilde{F}^{\textrm{em}\,\mu\nu}+  Z_{\mu\nu}\tilde{Z}^{\mu\nu} \right)\\
&=\frac{ P}{f}\,\frac{1}{32\,\pi^2}\left(2g_2^2 W^{+}_{\mu\nu}\tilde{W}^{{-}\,\mu\nu} + g_2^2 Z_{\mu\nu}\tilde{Z}^{\mu\nu}+ \frac{e^2}{\sin^2\theta_W} F^{\textrm{em}}_{\mu\nu}\tilde{F}^{\textrm{em}\,\mu\nu} \right)
}
where $Y_1$ is the properly normalized U(1) gauge field. Note that $g'=Cg_1, s_W^2=g^{'2}/G^2=C^2g_1^2/ (C^2g_1^2+g_2^2)=1
/(1+1/C^2)$, and ${c}^0_{a\gamma\gamma}=\frac{1}{\sin^2\theta_W}$. Note that $C^2=\frac53$ in the SU(5) model. Here, we used
 \dis{
 &W^3_\mu=\cos\theta_W Z_\mu +\sin\theta_W A_\mu,\\
 &Y_\mu=-\sin\theta_W Z_\mu +\cos\theta_W A_\mu ,\\
 &c_W=\cos\theta_W=\frac{g_2}{\sqrt{g_2^2+g^{\prime\,2}}},\\
 &s_W=\sin\theta_W=\frac{g'}{\sqrt{g_2^2+g^{\prime\,2}}}.
 }
Therefore, we obtain the following coupling for the anomalous case,
 \dis{
c_{a\gamma\gamma}= \frac{1-2\sin^2\theta_W}{\sin^2\theta_W}
 }
where we used $\frac{m_u}{m_d}=\frac12$.

   %%%%%%%%%%% 
\begin{table}[t!]
\begin{center}
   \begin{tabular}{cr|ccrl|lrl}
 &KSVZ~ &&&~DFSZ~&&&Superstring  & \quad\quad\quad Comments\\ 
$Q_{\rm em}$ &$c_{a\gamma\gamma}$ &  $x$&$q^c$-$e_L$ pair &$c_{a\gamma\gamma}$&
$c_{a\gamma\gamma}$&&$c_{a\gamma\gamma}$
&~~ \\[0.3em]  \hline &&&&&&&&\\[-1em]
$0$ &$-2$~  &~any $x$~&$(d^c,e)$   &$\frac{2}{3}$~&~~$\frac{2}{3}$ &~arXiv:1405.6175~& $\frac{2}{3}$&~Anomalous U(1) as U(1)$_{\rm PQ}$
\\[0.3em]
$\pm\frac13$ &$-\frac43$~ &~any $x$~&$(u^c,e)$   &$-\frac{4}{3}$~&~~$\frac{2}{3}$ &~hep-ph/0612107~& $-\frac{1}{3}$&~Approximate U(1)$_{\rm PQ}$
\\[0.3em]
$\pm\frac23$ &~~$\frac23$~ &~ ~&  &Without &~~GUTs or&~This paper ~& $\ge \frac{2}{3}$&~Anomalous U(1) as U(1)$_{\rm PQ}$
\\[0.3em]
$\pm 1$ &$4$~ &~ ~&  &SUSY&  ~~SUSY & & & $~c_{a\gamma\gamma}=(1-2\sin^2\theta_W)/\sin^2\theta_W$
\\[0.3em]
$(m,m)$ &$-\frac13$~ &~ ~&  & $H_d$ or  $H_u^*$&  ~~$H_{d}$ or  $H_u$& && ~~ with $m_u/m_d=0.5$.
\\
   \end{tabular}
   \end{center}
\caption{The axion-photon-photon coupling for several invisible axion models. The  third row  in superstring corresponds to the exact global symmetry U(1)$_{\rm PQ}=$U(1)$_{\rm anom}$ in the superpotential $W$. The MI axion with $f_a>10^{16\,}\gev$ \cite{ChoiKim85} has the same value as that of the KSVZ axion with $Q_{\rm em}=0$. The non-SUSY DFSZ models have a fine-tuning problem. One related cosmological problem even with SUSY was pointed out in \cite{Dreiner14}. }\label{tab:cagg}
\end{table}   
 %%%%%%%%%%
 
The axion-photon-photon couplings for invisible axions are summarized in Table \ref{tab:cagg}. In the DFSZ columns, the case with $H_u^*$ corresponds to that the $Q_{\rm em}=-1$ leptons obtain mass by the coupling $f_{ij} (\tilde{H}_u^T \bar{e}^i_R\ell^j_L)$ where $\tilde{H}_u=i\sigma_2 H_u^*, $ and $\ell^j_L=(\nu_j, e_j)_L^T$. In GUTs, both $d^c$ or $u^c$ has the same PQ charge as that of $\ell_i$ and $c_{a\gamma\gamma}$ are the same. With SUSY, holonomy forbids the coupling of $H_u^*$ to $\ell_i$ and only the coupling of $H_d$ to $\ell_i$ is allowed.

%%%%%%%%%%%%%%%%%%%%
\begin{figure}[!t]
\begin{center}
\includegraphics[width=0.55\linewidth]{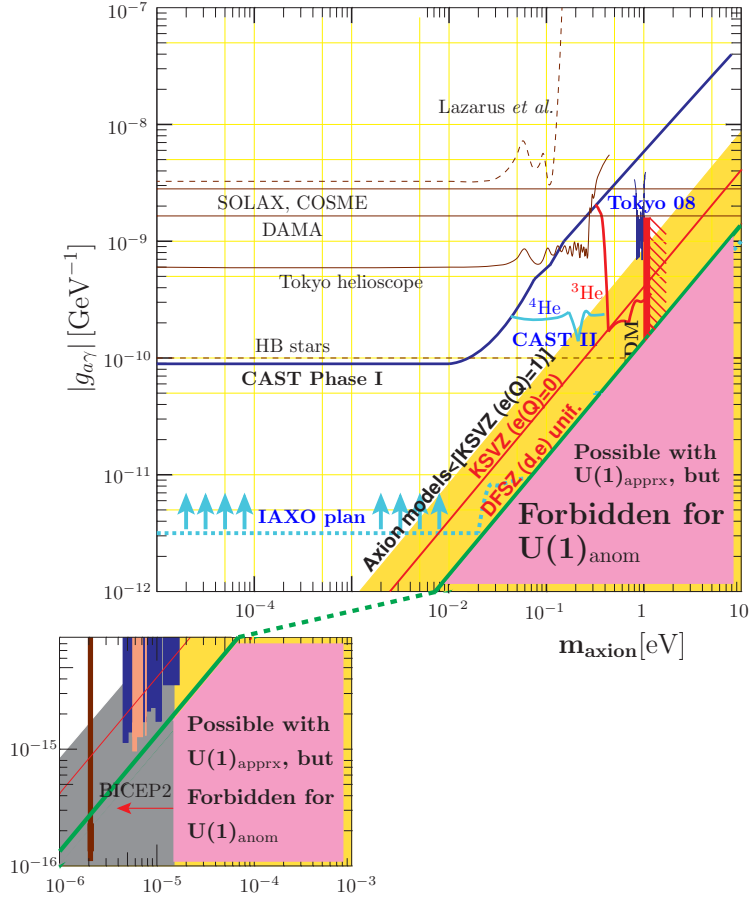}
\end{center}
\caption{The allowed parameter space of $g_{a\gamma}\, [\gev^{-1}]= 1.57\cdot 10^{-10}\,c_{a\gamma\gamma}$ vs. axion mass. The lavender part is not allowed if the U(1)$_{\rm PQ}$ is the anomalous U(1)$_{\rm anom}$. However, it can be allowed for some approximate U(1)$_{\rm PQ}$, as shown in Ref. \cite{ChoiKimIW07} for the flipped-SU(5) model of Ref. \cite{KimKyae07}.  } \label{fig:Uanom}
\end{figure}
%%%%%%%%%%%%%%%%%

%%%%%%%%%%%%%%
\subsection{The weak mixing angle in GUTs with extra U(1)'s}

If the electromagnetic charge operator is embedded in a simple GUT group SU($N$), the charge operator on the fundamental representation is a traceless matrix, 
\dis{
Q_{\rm em}({\bf N})={\rm diag.\,}( a, a, a, 0, -1, b_6, \cdots),~~3a-1+\sum_i b_i=0. \label{eq:QinSUN}
}
In the Georgi-Glashow (GG) SU(5) model \cite{GG74}, $a=\frac13$ and $b_i=0$. If $Q_{\rm em}$ is completed by the simple SU($N$) generators, the information on the fundamental representation is enough since the other higher dimensiomnal representations can be constructed in terms of direct products of fundamentals.  At the GUT scale three gauge couplings are the same, and the mixing angle is defined as $\sin\theta_W= e/g_2$. For properly normalized generators $Q_1$ and $T_3$ in the fundamental representation, the trace is $\frac12$, and we have
\dis{
\textrm{Tr\,}(eQ_{\rm em})^2=\textrm{Tr\,}(g_1 Q_1)^2=\textrm{Tr\,}(g_2 T_3)^2,
}
where $g_1=g_2$ at the GUT scale. Thus, we obtain
\dis{
\sin^2\theta_W=\frac{e^2}{g_2^2}=\frac{\textrm{Tr\,}(T_3)^2}{\textrm{Tr\,}(Q_{\rm em})^2}. \label{eq:WeakMixAngle}
}
For the SU(5) model, it is $(1/2)/(4/3)=3/8$. For the electromagnetic charge (\ref{eq:QinSUN}) in SU($N$), the mixing angle is
\dis{
\sin^2\theta_W =\frac{1/2}{3a^2+1+\sum_i b_i^2}. 
}
The SU(7) model of Ref. \cite{KimPRL80} gives $\sin^2\theta_W=3/20$.

If electromagnetically neutral $\GSM$  singlets are added to the fifteen chiral fields of SU(5), the weak mixing angle presented in Eq. (\ref{eq:WeakMixAngle}) remains the same. We can present the following general statement. Suppose that a GUT group breaks at one scale  $\Mg$ and matter fields breaks down to 45 chiral fields of the SM plus $\GSM$ singlets,
\dis{
3\big\{q_L,u^c_L,d^c_L,\ell_L,\nu_L,e_L,e^c_L \big\}+\textrm{singlets,~~at ~}\Mg.\label{eq:ThreeFSM}
}
Then, Eq. (\ref{eq:WeakMixAngle}) can be applied. Therefore, the SO(10) GUT has the weak mixing angle $\sin^2\theta_W=\frac38$. It does not depend on how the symmetry breaking chain takes, through the GG SU(5) or through the flipped SU(5) \cite{Barr82Flip,KimKyae07}, because there is only one scale $\Mg$. In the flipped-SU(5), there are three fermionic representations, ${\bf 10}_{+1/5},\overline{\bf 5}_{-3/5},$ and ${\bf 1}_{+1}$. If we consider all representations, Eq.  (\ref{eq:WeakMixAngle})  is still applicable. 

 However, if there are two scales for the symmetry breaking pattern such as $\textrm{SO(10)}\to \textrm{flipped-SU(5)}\to $\,SM, the weak mixing angle at the lower GUT scale has a logarthmic correction because U(1)$_{\rm em}$ is composed of  two U(1)'s. For the electromagnetic charge operator   composed of two  U(1) couplings, \ie $e^{2}_{N}$ from SU($N$) part and $e^2_{(1')}$ from U(1)$'$ part,  the electromagnetic charge is given by
\dis{
\frac{1}{e^2}=\frac{1}{e^{2}_{N}} +\frac{1}{e^{2}_{(1')}} . \label{eq:TwoCoupls}
}
If a vectorlike representation of the form ${\bf 5}_{-a}+\overline{\bf 5}_{a}$ is present in the model, then Eq.  (\ref{eq:WeakMixAngle})  is still applicable \cite{KimPLB14}. But, if ${\bf 5}$ or $\overline{\bf 5}$ does not appear in the anomaly-free combination as that from {\bf 16}, that fundamental representation cannot be used in Eq.  (\ref{eq:WeakMixAngle}). The Higgs ${\bf 5}_{-2/5}$ and $\overline{\bf 5}_{2/5}$ in the flipped-SU(5) gives $\sin^2\theta_W=\frac38$ via Eq. (\ref{eq:TwoCoupls}).

{  
The definition of $c_3$ and $c_{a\gamma\gamma}^0$ given in Eq. (\ref{eq:Defc3cem}) dictates that $f_a$ is the vacuum expectation value $\langle\sigma\rangle$ divided by the domain wall number $N_{\rm DW}$. $c_{a\gamma\gamma}^0$ is defined relative to $c_3$ with $c_3$ taking into account $N_{\rm DW}$. For three fundamental representations, $c_3$ is three times that of one fundamemental, and also  $c_{a\gamma\gamma}^0$  is three times that of one fundamemental. The color coupling defines $f_a$ as $\langle\sigma\rangle/N_{\rm DW}$. Thus,  $c_{a\gamma\gamma}^0$ is defined relative to $c_3$, \ie $c_{a\gamma\gamma}^0= {\textrm{Tr\,}(Q_{\rm em})^2}/  {\textrm{Tr\,}(F_3)^2}$ where $F_3$ is one generator of color gauge group SU(3)$_c$. For a fundamental representation, ${\textrm{Tr\,}(F_3)^2}={\textrm{Tr\,}(T_3)^2}$ where  $T_3$ is one generator of weak gauge group SU(2)$_W$, and we obtain
\dis{
c_{a\gamma\gamma}^0=\frac{\textrm{Tr\,}(Q_{\rm em})^2}{\textrm{Tr\,}(T_3)^2}=\frac{1}{\sin^2\theta_W},
}
if one fundamental representation is enough to calculate $c_{a\gamma\gamma}^0$.
Because the SM, represented in Eq. (\ref{eq:ThreeFSM}), has the contribution from  three families
\dis{
\frac{\textrm{Tr\,}(Q_{\rm em})^2}{\textrm{Tr\,}(F_3)^2}=\frac83,
}
the extra charged singlets will make this contribution larger. Therefore, GUTs predict
\dis{
c_{a\gamma\gamma}^0\ge \frac83.
}
Thus, we have the excluded region for the case of anomalous U(1) being  U(1)$_{\rm PQ}$ in Fig. \ref{fig:Uanom}. However, if U(1)$_{\rm PQ}$ is approximate as calculated in a string compactification   \cite{ChoiKimIW07}, this bound does not apply.
}

%%%%%%%%%%%%%%
\section{A simple calculation of $c_{a\gamma\gamma}^0$ from quantum numbers}

In this section, we show a simple method for calculating the entries in the DFSZ models in Table \ref{tab:cagg}. Let the invisible axion is housed in the complex singlet $\sigma$ \cite{KSVZ1}. The DFSZ model connects the PQ charges of $H_u$ and $H_d$ to that of $\sigma$. One possible connection is 
\dis{
H_uH_d\, \sigma^2,
}
and the PQ charge $\Gamma$ of $\sigma$ is assigned to be +1. The mass terms of the up- and down- type quarks are
\dis{
H_u^\dagger \bar{u}_R q_L,~H_d^\dagger \bar{d}_R q_L,
}
where $q_L$ and $\ell_L$ are SU(2)$_W$ doublets.

If the charged leptons obtain mass by $H_d$ via $H_d^\dagger \bar{e}_{ R} \ell_{ L} $, we can assign the charges of $H_u, q_L,\ell_L$, and $ u_R$ zero. Then, $H_d$ carries $-2$ units of the charge. The charges of $d_R$ and $e_R$ are +2. Certainly, this definition is free of gauge charges since fermions $d_R$ and $e_R$, having different SM gauge charges, have the same charge. Namely, these charges do not contain gauge charges. So, $-2$ is wholely the global PQ charge. Thus, $\Gamma-Q_{\rm em}-Q_{\rm em}$ anomaly is proportional to $+2e^2 [3(-1/3)^2+(-1)^2]=\frac{8e^2}{3}$, which is used in the table for $(d^c,e)$ unification.

If the charged leptons obtain mass by $H_u$ via $\tilde{H}_u^\dagger \bar{e}_{ R} \ell_{ L} $, we can assign the PQ charges of $H_d, q_L,\ell_L$, and $ d_R$ zero. Since $H_u$ carries $-2$ units of the PQ charge, the PQ charge of $u_R$ is +2, and the PQ charge of $e_R$ is $-2$. Thus, $\Gamma-Q_{\rm em}-Q_{\rm em}$ anomaly is proportional to $+2e^2 [3(+2/3)^2-(-1)^2]=\frac{2e^2}{3}$, which is used in the table for $(u^c,e)$ unification. In SUSY models, $H_u$ cannot be used for the electron mass due to the holomorphic condition and the weak hypercharge.
  
%%%%%%%%%%%%%%%%%%%%%%%%%%%%%%%%%%%%%%%%%%%%%%%%%%%%%%%%%%%%%%%%%%%%
\section{Conclusion}
 
For the exact global symmetry U(1)$_{\rm PQ}$ from string compactification, we obtained the lower bound, $\frac83-c_{a\gamma\gamma}^{\rm ch\,br}$, for the axion--photon--photon coupling $c_{a\gamma\gamma}$, where $c_{a\gamma\gamma}^{\rm ch\,br}\simeq 2$. This bound is free from the gravity obstruction of global symmetries. However, if  U(1)$_{\rm PQ}$ is approximate, this bound does not apply.

 %%%%%%%%%%%%%%%%%%%% 
\acknowledgments{This work has evolved from an earlier discussion with Peter Nilles and Patrick Vaudrevange on the related topic. I thank Bethe Center for Theoretical Physics, for the invitation to the Bethe Forum on ``Axions and the Low Energy Frontier'' (7--18 March 2016), where this work has been finished. 
This work is supported in part by the National Research Foundation (NRF) grant funded by the Korean Government (MEST) (NRF-2015R1D1A1A01058449) and  the IBS (IBS-R017-D1-2016-a00).}
\vskip 0.3cm

%%%%%%%%%%%%%%%%%%%%%%%%%%%%%%%%%%%%%%%%%%%%%%%%%%%%%%%%%%%%%%%%%%%%%%%%%%


\begin{thebibliography}{99}
\def\prp#1#2#3{{Phys.\,Rep.}  {\bf #1}  (#3) #2}
\def\rmp#1#2#3{{Rev. Mod. Phys.}  {\bf #1} (#3) #2}
\def\npb#1#2#3{{ Nucl.\,Phys.\,B}   {\bf #1}  (#3) #2}
\def\plb#1#2#3{{Phys.\,Lett.\,B}   {\bf #1}  (#3) #2}
\def\prd#1#2#3{{Phys.\,Rev.\,D}   {\bf #1}  (#3) #2}
\def\prl#1#2#3{{Phys.\,Rev.\,Lett.}   {\bf #1} (#3) #2}
\def\jhep#1#2#3{{JHEP}   {\bf #1} (#3) #2}
\def\jcap#1#2#3{{JCAP}   {\bf #1}  (#3) #2}
\def\zp#1#2#3{{Z.\,Phys.}   {\bf #1} (#3) #2}
\def\njp#1#2#3{{New\,J.\,Phys.}   {\bf #1}  (#3) #2}
\def\epjc#1#2#3{{Euro.\,Phys.\,J.\,C}    {\bf #1} (#3) #2}
\def\frp#1#2#3{{Front.\,Phys.}    {\bf #1}  (#3) #2}
\def\jpg#1#2#3{{J.\,Phys.\,G}   {\bf #1}  (#3) #2}
\def\ijmpd#1#2#3{{Int.\,J.\,Mod.\,Phys.\,D}   {\bf #1} (#3) #2}
\def\mpla#1#2#3{{Mod.\,Phys.\,Lett.\,A}   {\bf #1} (#3) #2}
\def\apj#1#2#3{{Astrophys.\,J.}    {\bf #1}  (#3) #2}
\def\nat#1#2#3{{Nature}    {\bf #1} (#3) #2}
\def\sjnp#1#2#3{{Sov.\,J.\,Nucl.\,Phys.}   {\bf #1} (#3) #2}
\def\apj#1#2#3{{Astrophys.\,J.}   {\bf #1}  (#3) #2}
\def\mnra#1#2#3{{Mon.\,Not.\,Roy.\,Astron.\,Soc.}    {\bf #1} (#3) #2}
\def\jetpl#1#2#3{{JETP\,Lett.}   {\bf #1}  (#3) #2}
\def\pthp#1#2#3{{Prog.\,Theor.\,Phys.}    {\bf #1} (#3) #2}
\def\jkps#1#2#3{{J.\,Korean\,Phys.\,Soc.}   {\bf #1} (#3) #2}
\def\dum#1#2#3{{\bf #1} (#3) #2}

\def\ibid#1#2#3{{\it ibid.}   {\bf #1} (#3) #2}
\def\err#1#2#3{{\bf #1}  (#3) #2\,(E)}   


\bibitem{KSVZ1}  J.E. Kim, \emph{Weak interaction singlet and strong CP invariance}, \prl{43}{103}{1979} [doi: 10.1103/PhysRevLett.43.103].

\bibitem{KSVZ2}  M.A. Shifman, V.I. Vainshtein, V.I. Zakharov, \emph{Can confinement ensure natural CP invariance of strong interactions?}, \npb{166}{493}{1980} [doi:10.1016/0550-3213(80)90209-6].

\bibitem{DFSZ} M. Dine, W. Fischler and M. Srednicki, \emph{A simple solution to the strong CP problem with a harmless axion}, \plb{104}{199}{1981} [doi:10.1016/0370-2693(81)90590-6]; A. P. Zhitnitsky, \emph{On possible suppression of the axion hadron interactions  (in Russian)}, Sov. J. Nucl. Phys. {\bf 31}, 260 (1980), Yad. Fiz. {\bf 31} (1980) 497.

\bibitem{SikivieDW} P. Sikivie, \emph{Of axions, domain walls and the early Universe}, \prl{48}{1156}{1982} [doi: 10.1103/PhysRevLett.48.1156].

\bibitem{KimRMP10} J. E. Kim and G. Carosi, \emph{Axions and the strong CP problem}, \rmp{82}{557}{2010} [arXiv: 0807.3125[hep-ph]].

\bibitem{ManoharPDG} A.V. Manohar and C.T. Sachrajda, \emph{Quark masses}, in K. Olive \etal (PDG collaboration), Chin. J. Phys. {\bf C\,38} (2015) 090001, p.725.

\bibitem{KimPRP87} J.E. Kim, \emph{Light pseudoscalars, particle physics, and cosmology}, \prp{150}{1}{1987}.

\bibitem{KimPRD98} J.E. Kim, \emph{Constraints on very light axions from cavity experiments}, \prd{58}{055006}{1998} [arXiv:hep-ph/98]. 

\bibitem{Kaplan85}
D.B. Kaplan, \emph{Opening the axion window}, \npb{260}{215}{1985} [doi:10.1016/0550-3213(85)90319-0]; \\
M. Srednicki, \emph{Axion couplings to matter: (I). CP-conserving parts}, \npb{260}{689}{1985} [do:10.1016/0550-3213(85)90054-9].  

\bibitem{PQ77} R. D. Peccei and H. R. Quinn, \emph{CP conservation in the presence of instantons},  \prl{38}{1440}{1977} [doi: 10.1103/PhysRevLett.38.1440].

\bibitem{ADMX} http://depts.washington.edu/admx/about-us.shtml

\bibitem{BCM14} J.E. Kim, Y.K. Semertzidis, and S. Tsujikawa, \emph{Bosonic coherent motions in the Universe}, \frp{2}{60}{2014} [arXiv:1409.2497 [hep-ph]].

\bibitem{BarrGr92}
S. M. Barr and D. Seckel, \emph{Planck scale corrections to axion models}, \prd{46}{539}{1992} [doi: 10.1103/PhysRevD.46.539];\\
M. Kamionkowski and J. March-Russell,
\emph{Planck scale physics and the Peccei-Quinn mechanism}, \plb{282}{137}{1992} [hep-th/9202003];\\
R. Holman, S. D. H. Hsu, T. W. Kephart, E. W. Kolb, R. Watkins, and L. M. Widrow, \emph{Solutions to the strong CP problem in a world with gravity}, \plb{282}{132}{1992} [hep-ph/9203206];\\
B. A. Dobrescu, \emph{The strong CP problem versus Planck scale physics}, \prd{55}{5826}{1997} [hep-ph/9609221].

\bibitem{Giddings87} S.B. Giddings and A. Strominger, \emph{Axion induced topology change in quantum gravity and string theory},
\npb{306}{890}{1988} [doi:10.1016/0550-3213(88)90446-4].

\bibitem{Coleman88} S.R. Coleman, \emph{Why there is nothing rather than something: A theory of the cosmological constant}, \npb{310}{643}{1988} [doi:10.1016/0550-3213(88)90097-1].

\bibitem{StroWitten85}
A. Strominger and E. Witten, \emph{New manifolds for superstring compactification}, Commun. Math. Phys. {\bf 101} (1985) 341 [doi:10.1007/BF01216094].

\bibitem{DiscrGauge89} L.M. Krauss and F. Wilczek, \emph{Discrete gauge symmetry in continuum theories}, \prl{62}{1221}{1989} [doi: 10.1103/PhysRevLett.62.1221].

\bibitem{Ibanez92} L.E. Ibanez and G.G. Ross, \emph{Discrete gauge symmetries and the origin of baryon and lepton number conservation in supersymmetric versions of the standard model}, \plb{368}{3}{1992} [doi: 10.1016/0550-3213(92)90195-H].

\bibitem{BanksDine92} T. Banks and M. Dine, \emph{Note on discrete gauge anomalies }, \prd{45}{1424}{1992} [arXiv:hep-th/9109045].

\bibitem{Preskill91} J. Preskill, S. P. Trivedi, F. Wilczek, and M. B. Wise, \emph{Cosmology and broken discrete symmetry}, \npb{363}{207}{1991} [doi:10.1016/0550-3213(91)90241-O].

\bibitem{KimPRL13} J.E. Kim, \emph{Natural Higgs-flavor-democracy solution of the $\mu$ problem of supersymmetry and the QCD axion}, \prl{111}{031801}{2013} [arXiv:1303.1822 [hep-ph]].

\bibitem{KimPLB13} J.E. Kim, \emph{Abelian discrete symmetries $\Z_N$ and $\Z_{nR}$ from string orbifolds}, \plb{726}{450}{2013} [arXiv: 1308.0344[hep-th]].

\bibitem{Kobayashi07} An example is shown in, T. Kobayashi, H.P. Nilles, F. Fl\"oger, S. Raby, and M. Ratz, \emph{Stringy origin of non-Abelian discrete flavor symmetries}, \npb{768}{135}{2007} [arXiv:hep-ph/0611020].

\bibitem{WittenMD} E. Witten, \emph{Cosmic superstrings}, \plb{153}{243}{1985} [doi:10.1016/0370-2693(85)90540-4].

\bibitem{WenWitten} X.G. Wen and E. Witten, \emph{World sheet instantons and the Peccei-Quinn symmetry}, \plb{166}{397}{1986} [doi:10.1016/0370-2693(86)91587-X].

\bibitem{GreenSch84} M.B. Green and J. Schwarz, \emph{Anomaly cancellation in supersymmetric D=10 gauge theory and superstring theory}, \plb{149}{117}{1984} [doi:10.1016/0370-2693(84)91565-X].

\bibitem{WittenMI} E. Witten, \emph{Some properties of O(32) superstrings}, \plb{149}{351}{1984} [doi:10.1016/0370-2693(84)90422-2].

\bibitem{ChoiKim85} K. Choi and J. E. Kim, \emph{Harmful axions in superstring models}, \plb{154}{393}{1985} and \err{156}{452}{1985} [doi:10.1016/0370-2693(85)90416-2].

\bibitem{Pi84} S-Y. Pi, \emph{Inflation without tears}, \prl{52}{1725}{1984} [doi:10.1103/PhysRevLett.52.1725].

\bibitem{AnomUone} J.J. Atick, L. Dixon, and A. Sen, \emph{String calculation of Fayet-Iliopoulos $d$ terms in arbitrary supersymmetric compactifications}, \npb{292}{109}{1987} [doi:10.1016/0550-3213(87)90639-0];\\
M. Dine, I. Ichinose, and N. Seiberg, \emph{F terms and $d$ terms in string theory}, \npb{293}{253}{1987} [doi:10.1016/0550-3213(87)90072-1].

\bibitem{KimPLB88} J. E. Kim, \emph{The strong  CP problem in orbifold compactifications and an SU(3)$\times$SU(2) $\times$U(1)$^n$ model}, \plb{207}{434}{1988} [doi:10.1016/0370-2693(88)90678-8].

\bibitem{KimPLB14} J. E. Kim, \emph{Calculation of axion--photon--photon coupling in string theory}, \plb{735}{95}{2014}  and \err{741}{327}{2014} [arXiv:1405.6175 [hep-ph]].

\bibitem{Dreiner14} H.K. Dreiner, F. Staub, and L. Ubaldi, \emph{From the unification scale to the weak scale: A self consistent supersymmetric Dine-Fischler-Srednicki-Zhitnitsky axion model},  \prd{90}{055016}{2014} [arXiv:1402.5977 [hep-ph]].

\bibitem{ChoiKimIW07}  K.-S. Choi, I.-W. Kim and J. E. Kim, \jhep{0703}{116}{2007} [arXiv:hep-ph/0612107].

\bibitem{KimKyae07}  
J.E. Kim and B. Kyae, \emph{Flipped SU(5) from $\Z_{12-I}$ orbifold with Wilson line}, \npb{770}{47}{2007}  [arXiv:hep-th/0608086].  

\bibitem{GG74} H. Georgi and S.L. Glashow, \emph{Unity of all elementary particle forces}, \prl{32}{438}{1974} [doi: 10.1103/PhysRevLett.32.438].

\bibitem{KimPRL80} J.E. Kim, \emph{Model of flavor unity}, \prl{45}{1916}{1980} [doi:10.1103/PhysRevLett.45.1916].

\bibitem{Uranga99} L.E. Ibanez, R. Rabadan, and A.M. Uranga,
\emph{Anomalous U(1)'s in type I and type IIB D = 4, N=1 string vacua},  \npb{542}{112}{1999} [arXiv:hep-th/9808139].

\bibitem{Barr82Flip} S.M. Barr, \emph{A new symmetry breaking pattern for SO(10) and proton decay}, \plb{112}{219}{1982} [doi:10.1016/0370-
2693(82)90966-2];\\
 J-P. Derendinger, J.E. Kim, and D.V. Nanopoulos, \emph{Anti-SU(5)}, \plb{139}{170}{1984} [doi:10.1016/0370-
2693(84)91238-3];\\
 I. Antoniadis, J.R. Ellis, J.S. Hagelin, and D.V. Nanopoulos, \emph{The flipped SU(5)$\times$U(1) string
model revamped}, \plb{231}{65}{1989} [doi:10.1016/0370-2693(89)90115-9].

\end{thebibliography}
\end{document}